\documentclass[journal=jacsat,manuscript=article,doi=true]{achemso}
\setkeys{acs}{doi=true}

\usepackage[version=3]{mhchem} 
\usepackage[utf8]{inputenc}
\usepackage{graphicx}
\usepackage{dcolumn}
\usepackage{bm}
\usepackage{blindtext}
\usepackage{fancyvrb}
\usepackage[normalem]{ulem}
\usepackage{arydshln}

\usepackage{siunitx}
\usepackage{algorithm}
\usepackage{algpseudocode}
\usepackage{graphicx}
\usepackage{subcaption}

\usepackage[justification=centering]{caption}

\usepackage{tikz}
\usepackage{ulem}
\usetikzlibrary{shapes,arrows}
\usepackage{multirow}
\usepackage{amsmath}

\DeclareSIUnit\angstrom{\text {Å}}

\mciteErrorOnUnknownfalse

\author{Côme Cattin}
\affiliation{Sorbonne Université, Laboratoire de Chimie Théorique, UMR 7616 CNRS, 75005 Paris, France}

\author{Thomas Plé}

\affiliation{Sorbonne Université, Laboratoire de Chimie Théorique, UMR 7616 CNRS, 75005 Paris, France}
\email{thomas.ple@sorbonne-universite.fr}
\author{Olivier Adjoua}
\affiliation{Sorbonne Université, Laboratoire de Chimie Théorique, UMR 7616 CNRS, 75005 Paris, France}
\author{Nicolaï Gouraud}
\affiliation{Qubit Pharmaceuticals, Advanced Research Department, 75014 Paris, France}
\author{Louis Lagardère}
\affiliation{Sorbonne Université, Laboratoire de Chimie Théorique, UMR 7616 CNRS, 75005 Paris, France}

\author{Jean-Philip Piquemal}
\affiliation{Sorbonne Université, Laboratoire de Chimie Théorique, UMR 7616 CNRS, 75005 Paris, France}
\email{jean-philip.piquemal@sorbonne-universite.fr}

\title{Accelerating Molecular Dynamics Simulations with Foundation Neural Network Models using Multiple Time-Step and Distillation}


\begin{document}

\begin{abstract}
We present a distilled multi-time-step (DMTS) strategy to accelerate molecular dynamics simulations using foundation neural network models. DMTS uses a dual-level neural network where the target accurate potential is coupled to a simpler but faster model obtained via a distillation process. The 3.5 \AA-cutoff distilled model is sufficient to capture the fast-varying forces, i.e., mainly bonded interactions, from the accurate potential allowing its use in a reversible reference system propagator algorithms (RESPA)-like formalism. The approach conserves accuracy, preserving both static and dynamical properties, while enabling to evaluate the costly model only every 3 to 6 fs depending on the system. Consequently, large simulation speedups over standard 1 fs integration are observed: nearly 4-fold in homogeneous systems and 3-fold in large solvated proteins through leveraging active learning for enhanced stability. Such a strategy is applicable to any neural network potential and reduces their performance gap with classical force fields.

\end{abstract}

%%%%%%%% Intoduction

Neural Network potentials (NNPs) ~\cite{behler2007generalized,bartok2010gaussian,chmiela2018towards,shakouri2017accurate,smith2017ani,schutt2017schnet,unke2021spookynet,gasteiger2021gemnet,musaelian2022learning,grisafi2019incorporating,ANI2X,AIMNET,grisafi2021multi,gems,tkatchenko_fragment} have emerged as powerful tools to perform molecular dynamics (MD) simulations, offering near-quantum mechanical accuracy at a fraction of the computational cost of \textit{ab initio} methods. Presently, such models can be generalized up to Foundation Models  covering complete area of applications of molecular dynamics ranging from Material Science to Chemistry, Biology and Drug Design \cite{batatia2022mace,kovacs2023mace,ple2023fennix,Ple-2025,benali2025}.
By learning the potential energy surface from reference quantum chemistry data, these models enable simulations of large and complex molecular systems with significantly improved fidelity compared to classical force fields\cite{MACKERELL200591}.
However, this increase in accuracy comes at a cost: NNPs are substantially more expensive to evaluate than traditional empirical potentials. This limits their applicability in term of system sizes and long-timescale simulations.

One of the primary bottlenecks of performing MD with ML potentials lies in the time integration itself. The equations of motion must be solved with a small time step - typically in the order of \SI{0.5} to \SI{1} femtosecond - to resolve high-frequency motions such as bond vibrations. With computationally intensive ML models, this results in a high number of expensive force evaluations, further amplifying the overall simulation cost.

Multi-Time-Step (MTS) integrators \cite{RESPA,zhou2001efficient,lagardere2019pushing} offer a well-established strategy to address this challenge. Originally introduced in the context of classical molecular simulations, MTS methods such as RESPA (Reference System Propagator Algorithm) \cite{RESPA} exploit the separation of timescales between different components of the forces to reduce the number of expensive evaluations. By integrating fast-changing forces with a small time step and slow-changing ones with a larger time step, significant performance gains can be achieved without compromising accuracy.
Having proven their usefulness for classical force fields simulations, MTS schemes have been shown to be also applicable to Ab Initio MD (AIMD)\cite{UrsulaMTS}.  In the context of machine learning molecular simulations, few works were reported in that direction\cite{inizan2022scalable,fu2023learning,QiangMTS,Ursula25}. Thus, MTS schemes are not presently part of today’s NNPs toolkit. Indeed, in their case, force decomposition is not naturally given by physical interactions and naive MTS implementations can be
plagued by various sources of instabilities, for example resonances\cite{Skeelresonance1, Skeelresonance2}. Still, MTS represents an important opportunity for foundation models as the use of multiple neural networks of differing complexity and inference cost could enable efficient MTS schemes specifically tailored to ML potentials.

In this work, we propose the first global MTS strategy for neural network foundation models. As it relies on knowledge distillation~\cite{hinton2015distilling}, we term it DMTS for distilled multi-timestep. It is applicable to any type of systems and ensures stable long timescale simulations with consequent speedups. We leverage the RESPA formalism since it preserves simplecticity and time-reversibility~\cite{RESPA}. We integrate this approach into the FeNNol library \cite{ple2024fennol}, a neural network framework for molecular simulations that can be coupled to the scalable Deep-HP machine learning interface \cite{inizan2022scalable} present in the GPU-accelerated Tinker-HP molecular dynamics package \cite{lagardere2018tinker,adjoua2021tinker}. Our method associates a small, fast NNP obtained via distillation with the FeNNix-Bio1(M) foundation model \cite{Ple-2025}, leveraging the speed of the former to reduce the frequency of the expensive evaluations of the latter. We systematically evaluate the method using various systems including liquid water, solvated small molecules and proteins in the condensed phase, exploring different hyperparameter settings and model sizes in order to design the fastest setup while maintaining accuracy. Throughout, we compare the relative merits of both system-specific and general-purpose models as fast component for the DMTS scheme.
We also discuss the use of active learning to systematically increase DMTS's accuracy and performances.

Our results show that this hybrid RESPA-like scheme enables substantial computational speed-ups with limited loss of accuracy, offering a practical route to scalable and efficient Molecular Dynamics with foundation machine learning models and NNPs in general.

%%%%%%%% Methods

%%%%%%% Integration scheme
In this work, we use two neural network potentials of differing complexity to implement a MTS integration scheme. In this scheme, the dynamics of a cheaper NNP is integrated with a small time step ($\sim$1 fs) and is periodically corrected (every 2 to 6 steps depending on the system) by the force difference between the small NNP and a larger reference model, here the FeNNix-Bio1(M) model~\cite{Ple-2025}. This enables to recover the dynamics of the larger NNP without evaluating its forces at every time step, thus improving the computational efficiency. In particular, we employ the BAOAB-RESPA integration scheme~\cite{lagardere2019pushing} outlined in Algorithm~\ref{algo:baoab-respa}, that we implemented in the FeNNol library. For simulations in Tinker-HP, we use the implementation described in~\cite{lagardere2019pushing} and the Deep-HP interface~\cite{inizan2022scalable} for calling models.

\begin{algorithm}[htbp]
\caption{MTS Integration Step with FENNIX Force Splitting}
\label{algo:baoab-respa}
\begin{algorithmic}[1]
\If{$\text{first\_step}$}
\State $F_\text{small} \gets \text{FENNIX}_\text{small}(x)$ 
\State $F \gets \text{FENNIX}_\text{large}(x)$
\EndIf
\State $v \gets v + \dfrac{\Delta t}{2m} \cdot (F - F_\text{small})$
\For{$i = 1$ to $n_\text{slow}$}
    \State $v \gets v + \dfrac{\Delta t}{2m \cdot n_\text{slow}} \cdot F_\text{small}$
    \State $x \gets x + \dfrac{\Delta t}{2 \cdot n_\text{slow}} \cdot v$
    \State $v \gets \text{thermo}(v,\dfrac{\Delta t}{n_\text{slow}})$ \Comment{Apply thermostat}
    \State $x \gets x + \dfrac{\Delta t}{2 \cdot n_\text{slow}} \cdot v$
    \State $F_\text{small} \gets \text{FENNIX}_\text{small}(x)$
    \State $v \gets v + \dfrac{\Delta t}{2m \cdot n_\text{slow}} \cdot F_\text{small}$
\EndFor
\State $F \gets \text{FENNIX}_\text{large}(x)$
\State $v \gets v + \dfrac{\Delta t}{2m} \cdot (F - F_\text{small})$
\end{algorithmic}
\end{algorithm}

In Algorithm~\ref{algo:baoab-respa}, $x$ denotes the system's coordinates, $v$ its velocities, $\Delta t$ the outer time step, $m$ the mass and $n_\text{slow}$ the number of inner steps with time step $\Delta t/n_\text{slow}$. $\text{FENNIX}_\text{large}(x)$ denotes the reference FeNNix-Bio1(M) machine-learned force field evaluated at configuration $x$ and $\text{FENNIX}_\text{small}(x)$ denotes the cheaper model. We describe the neural networks architectures and the distillation strategies employed to train the $\text{FENNIX}_\text{small}$ model in the next sections.

%%%%%%%%%%% Neural Network Architectures

The reference model that we intend to accelerate corresponds to the FeNNix-Bio1(M) model, trained on a broad and diverse dataset. This model is based on a range-separated equivariant transformer architecture where close-range and long-range interactions are described with different spatial resolutions and dedicated attention heads. Its receptive field is \SI{11}{\angstrom} in total, via two message-passing interactions. Details about the full architecture are provided in ref.~\cite{Ple-2025}. 
The second, lighter-weight, model that is used in inner steps of the RESPA scheme uses the same base architecture but with reduced capacity and removes the long-range attention heads in order to enable faster inference and lower computational cost. Its receptive field is only \SI{3.5}{\angstrom} (one message-passing interaction), making it much more short-sighted and allowing it to focus mostly on fast-varying ``bonded'' forces.
The model hyperparameters that we used in the following numerical experiments are provided in Supporting Information.

%%%%%%%%%% Training procedures

We derive the $\text{FENNIX}_\text{small}$ model from the reference one via a knowledge distillation procedure~\cite{hinton2015distilling,gou2021knowledge} where the distilled model is trained on data labeled with the FeNNix-Bio1(M) model instead of DFT. This ensures that the forces used in the inner loop of the RESPA scheme are as close to the reference ones as possible, minimizing the correction necessary to recover the correct dynamics and thus the frequency of its application.
In practice, the small model comes in two flavors, either i) as an on-the-fly system-specific model or ii) as a generic model. We briefly describe here these two distillation strategies. Specific details about the training parameters can be found in Supporting Information.

\paragraph{System-specific model}

For each system, a reference dataset is generated by running a short MD simulation (less than a nanosecond) using the reference model. For proteins and, more generally, for large systems, a fragmentation strategy similar to the one proposed in ref.~\cite{tkatchenko_fragment} was employed to reduce the computational burden while retaining local structural information. 
We then evaluate the energies and forces of the collected frames with the reference model. The on-the-fly system-specific model is then trained on this dataset, resulting in a model about 10 times faster than FeNNix-Bio1(M) with our current setup.

\paragraph{Generic model}

In addition to system-specific distilled models, we also propose a generic fast model trained on a chemically diverse dataset, enabling broader applicability and faster deployment in new systems. To construct this transferable potential, we generate a training set by evaluating FeNNix-Bio1(M) on a subset of conformations from the SPICE2 dataset~\cite{eastman2023spice,eastman2024spice2}, which contains a wide variety of small organic molecules and biologically-relevant complexes.
The generic model is trained on the obtained dataset, resulting in a transferable model that captures general chemical patterns and can be reused across systems.
This generic model can be used directly in the inner loop of our DMTS scheme, or serve as an initialization point for further fine-tuning on a target system. This offers a compromise between generality and accuracy, especially when system-specific data is limited or a rapid deployment is desired.

%%%%%%%% Experiments

The results obtained with the DMTS scheme for each of the following experiments are compared to a single time step (STS) integrator using the FeNNix-Bio1(M) force field with a time step of \SI{1}{fs}.
The STS scheme employs a BAOAB Langevin integrator~\cite{leimkuhler2013rational,leimkuhler2013robust}. All simulations use a friction coefficient of \SI{1}{ps^{-1}} for the coupling to the Langevin thermostat.

\subsection{Bulk Water}
First, to investigate the robustness of our models and integrator, we performed a series of stability tests on a small water box containing 648 atoms.
Simulations were carried out using the system-specific model and the generic model, with external time steps ranging from \SI{2}{fs} to \SI{9}{fs}.
For each configuration, we monitored dynamical and thermodynamical observables, including diffusion coefficients, kinetic and potential energies, and temperature. 
Diffusion coefficients were computed using Tinker\cite{rackers2018tinker} with the Einstein formula.
The stability of the trajectory was evaluated over the full course of the \SI{2}{ns} simulation, with qualitative indicators reported in Table~\ref{tab:stability-tests}. Additional tests were performed with hydrogen mass repartitioning (HMR)~\cite{HMR}, enabling larger time steps.

\begin{table}
    \centering
    \begin{tabular}{l|c|c|c|c}
                 & \multicolumn{2}{c|}{Diffusion} & \multicolumn{2}{c}{Temperature}\\
                 & System & Generic & System & Generic\\
            \hline
      STS        &  \multicolumn{2}{c|}{2.21 $\pm$ 0.15} & \multicolumn{2}{c}{300.1$\pm$9.6}\\
      \SI{2}{fs} &  2.45 $\pm$ 0.17 & 2.06 $\pm$ 0.05 & 300.5$\pm$9.6 & 300.2$\pm$9.6 \\
      \SI{3}{fs} &  2.30 $\pm$ 0.06 & 2.66 $\pm$ 0.67 & 315.2$\pm$10.6 & 333.3$\pm$12.2\\
      \SI{3}{fs} HMR & 2.27 $\pm$ 0.14 & 2.05 $\pm$ 0.18 & 300.1$\pm$9.5 & 299.9$\pm$9.5\\
      \SI{4}{fs} HMR & 2.11 $\pm$ 0.19 & 2.04 $\pm$ 0.24 & 300.8$\pm$9.5 & 303.4$\pm$9.9\\
      \SI{5}{fs} HMR & 2.30 $\pm$ 0.19 & 2.16 $\pm$ 0.25 & 302.7$\pm$9.8 & 304.1$\pm$10.2\\
      \SI{6}{fs} HMR & 2.20 $\pm$ 0.30 & 2.64 $\pm$ 0.47 & 304.5$\pm$9.7 & 305.9$\pm$9.9\\
      \SI{7}{fs} HMR & NaN & 3.32 $\pm$ 0.04 & NaN & 327.5$\pm$11.1
    
    \end{tabular}
    \caption{
    Stability tests performed on a water box of 648 atoms using both the system-specific and generic models at various integration time steps, with and without hydrogen mass repartitioning (HMR).
    Diffusion coefficients (in $10^{-5}$ cm$^2$/s) and average temperatures (K) are reported with associated standard deviations.
    Values are averaged over \SI{2}{ns} trajectories.
    Instabilities prevented completion of simulations beyond \SI{3}{fs} without HMR and beyond \SI{7}{fs} with HMR.
}
    \label{tab:stability-tests}
\end{table}

At short external time step (\SI{2}{fs}), both specific and generic models produced stable trajectories with diffusion coefficients and temperatures close to the reference STS values.
At \SI{3}{fs}, simulations remained stable, though the generic and system-specific model displayed a marked increase in both diffusion coefficient and temperature, suggesting the onset of integration artifacts.

The introduction of HMR enables stable simulations up to \SI{6}{fs}, with dynamical and thermodynamical properties remaining within reasonable agreement between STS and DMTS integrator for both models.
However, at \SI{7}{fs} and beyond, instabilities systematically occurred, reflected by divergent values in Table~\ref{tab:stability-tests} or nonphysical increase in diffusion and temperature.
These results highlight that HMR provides a substantial extension of the accessible time step, effectively doubling the stability range compared to standard masses. 
Nevertheless, care must be taken to avoid excessive increases as instabilities rapidly emerge beyond \SI{6}{fs}. In practice, time steps of \SI{2}{fs} to \SI{3}{fs} without HMR, and up to 5-\SI{6}{fs} with HMR, appear to offer the most reliable compromise between computational efficiency, stability, and physical accuracy. 

\begin{figure}
    \centering
        \includegraphics[width=0.8\linewidth]{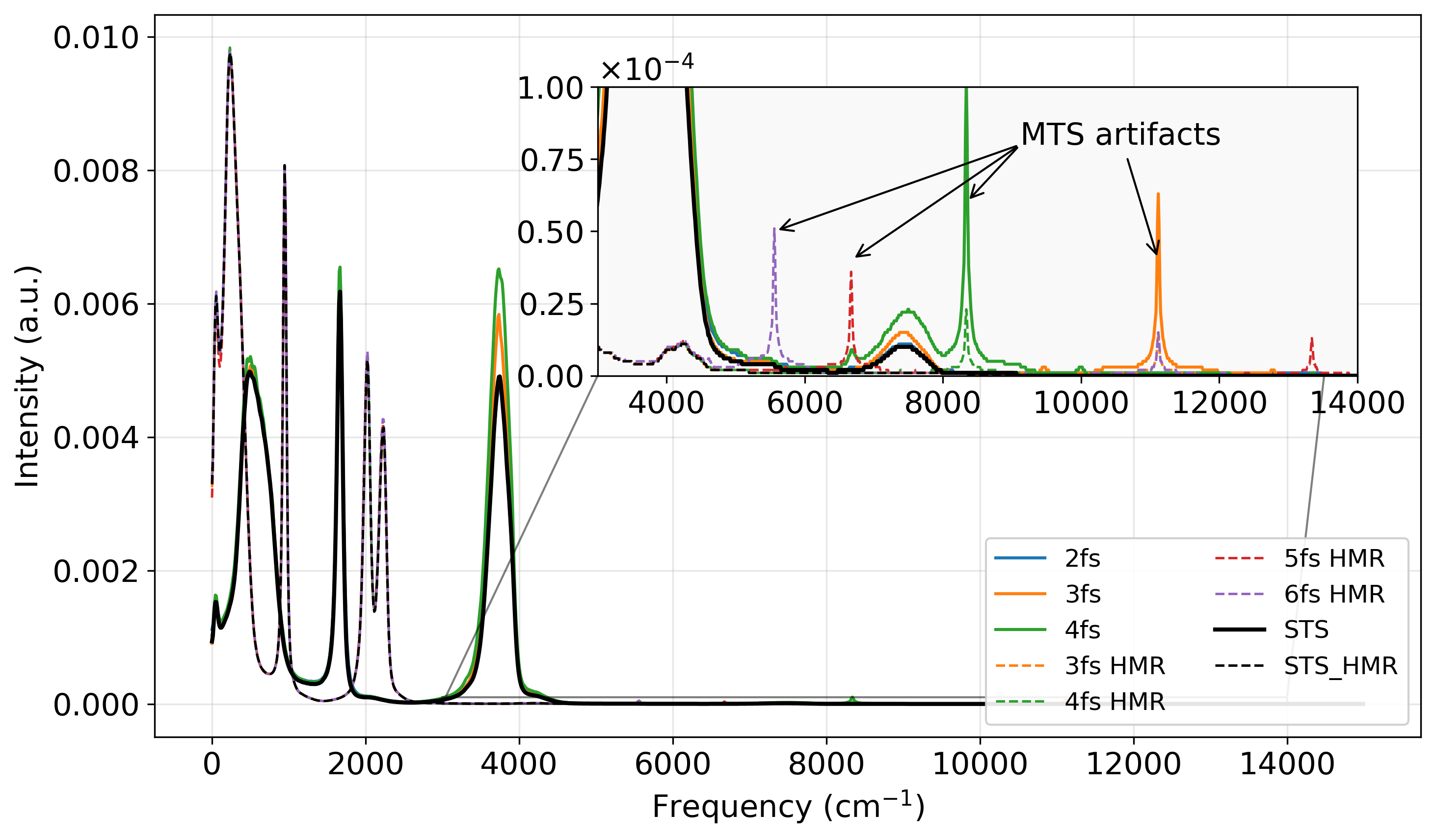}
    \caption{
        Average velocity autocorrelation spectra of hydrogen atoms for bulk water. 
        The reference single time step (STS) simulation are shown in black. 
        Distilled Multi time step (DMTS) simulations are shown for different integration time steps (2~fs, 3~fs, 4~fs, 5~fs, and 6~fs), 
        with corresponding heavy-mass repartitioning (HMR) variants represented by dashed lines of the same color.
        Inset zooms in high frequency regions containing MTS integration artifacts.
    }
    \label{fig:spectra_water}
\end{figure}

It is well documented that limits in the larger time step usable in the context of MTS schemes in molecular dynamics are related to the coupling of this largest time step with the highest-frequency vibrational modes involved in the system: the so-called resonance effects~\cite{Skeelresonance1,Skeelresonance2,GLE,isokin,isokinpol,isokin3,gouraudhmc,velocity_jumps}. A simple diagnostic can be made to assess these aspects by considering the velocity autocorrelation spectra in this setup. Figure \ref{fig:spectra_water} illustrates this in the context of bulk water where we see the artifacts produced by the periodic MTS force correction (annotated in the inset of Figure~\ref{fig:spectra_water}) progressively coupling with the overtone of the O-H stretching mode (around \SI{7500}{cm^{-1}} or \SI{4000}{cm^{-1}} with HMR) as the outer time step increases. This diagnostic confirms the limits of the outer time step to \SI{5}{fs} with hydrogen mass repartitioning and \SI{2}{}-\SI{3}{fs} without.

\begin{figure}
    \centering
    \includegraphics[width=0.53\linewidth]{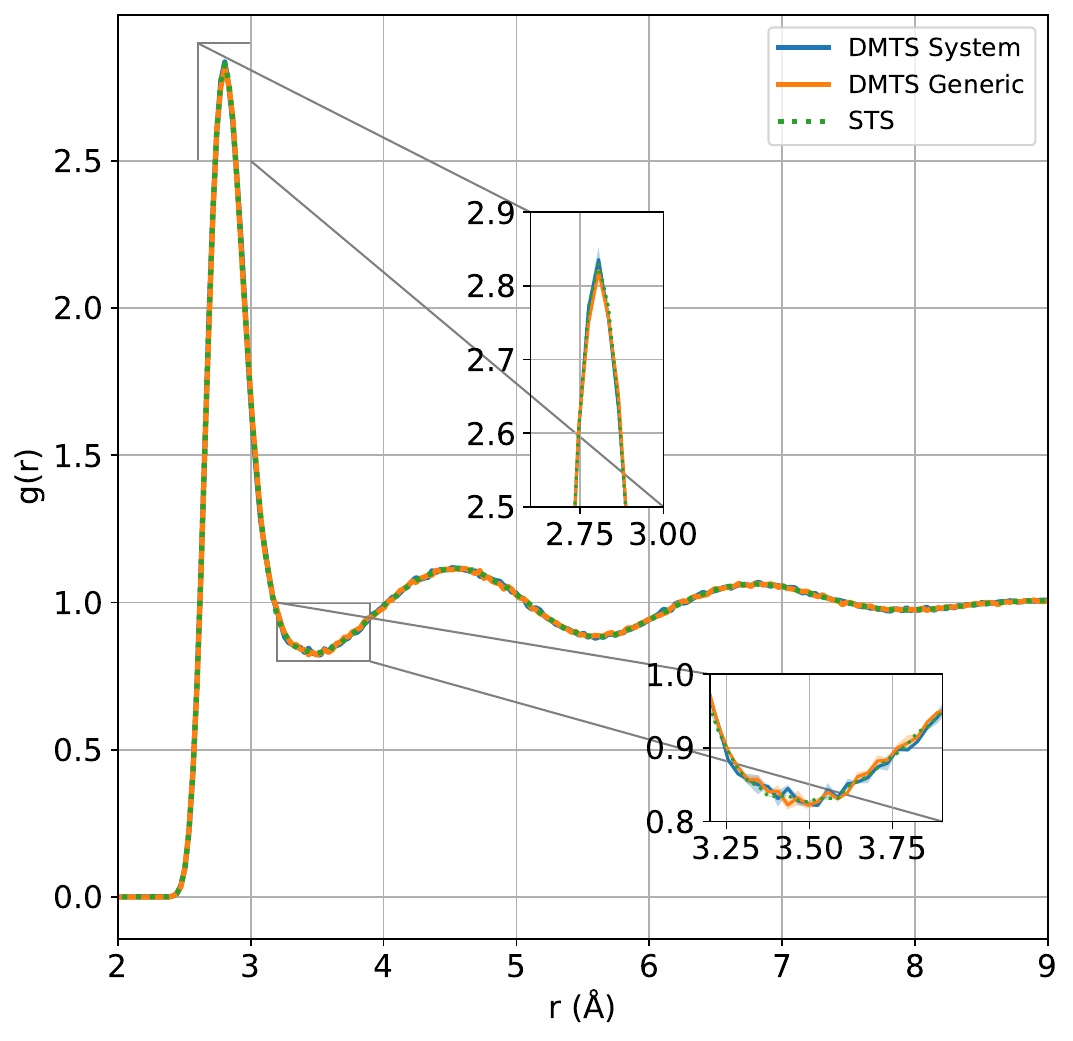}
    \caption{
        Radial distribution function, $g(r)$, as a function of the distance ($r$) in \SI{}{\text{Å}}.
        The solid blue curve corresponds to the DMTS simulation with the on-the-fly system-specific model and the orange one to DMTS with the small generic model. Both DMTS simulations used an inner time step of \SI{1}{fs} and an outer time step of \SI{5}{fs}. The dotted green curve corresponds to the reference STS simulation.
    }
    \label{fig:g_r_water}
\end{figure}

We further assess the robustness of our approach by computing the radial distribution function of a larger water box composed of 4800 atoms. As shown in Figure~\ref{fig:g_r_water}, the DMTS simulations (with both generic and system-specific models) with an outer time step of \SI{5}{fs} and HMR correctly reproduce the STS results within statistical uncertainties.
Overall, with this setup, we obtain a speedup of around 4 compared to STS with \SI{1}{fs} time step for bulk water simulations corresponding to an increase from \SI{6.59} to \SI{25.03}{ns/day} (See Table~\ref{tab:water-perf}).

\begin{table}
    \centering
    \begin{tabular}{l|c|c|c|c}
                 & \multicolumn{2}{|c}{Small} & \multicolumn{2}{|c}{Large}\\
                 & System & Generic & System & Generic\\
            \hline
      STS        & \multicolumn{2}{c}{38.07} & \multicolumn{2}{c}{6.59} \\
      \SI{2}{fs}     & 48.05  & 40.58 & 11.61 & 10.91 \\
      \SI{3}{fs}     & 67.81  & 57.57 & 16.26 & 15.26 \\
      \SI{4}{fs} HMR & 85.92  & 71.73 & 20.57 & 19.22 \\
      \SI{5}{fs} HMR & 98.69  & 84.85 & 25.03 & 22.81 \\
      \SI{6}{fs} HMR & 117.99 & 95.67 & 27.52 & 26.03 \\
    
    \end{tabular}
    \caption{
    Maximum performance on small (648 atoms) and large (4800 atoms) water boxes in \SI{}{ns/day} obtained on a single NVIDIA A100 GPU using both the system-specific and generic models at various integration time steps, with and without hydrogen mass repartitioning (HMR).
}
    \label{tab:water-perf}
\end{table}

\subsection{Solvated Molecules}

To further assess the accuracy of the proposed DMTS scheme, we turn to small solvated molecules. We observe stability limits similar to those for bulk water for the five molecules that we tested: ethanol, benzene, trimethylamine, diethylsulfide, and acetic acid. We then evaluated the hydration free energies (HFEs) of these (as well as water). Calculations were performed leveraging the alchemical Lambda-ABF~\cite{lagardere2024lambda} method, employing both system-specific models trained on-the-fly and a generic small foundation model. All simulations were carried out using the same DMTS integration scheme described above, with an inner time step of \SI{1}{fs} and an outer time step of \SI{4}{fs} and HMR. For benzene DMTS simulations using the generic model, we needed to reduce the outer time step to \SI{3}{fs} to avoid instabilities. This confirms that the system specific model is more robust than the generic one.

\begin{figure}
    \centering
    \includegraphics[width=0.5\linewidth]{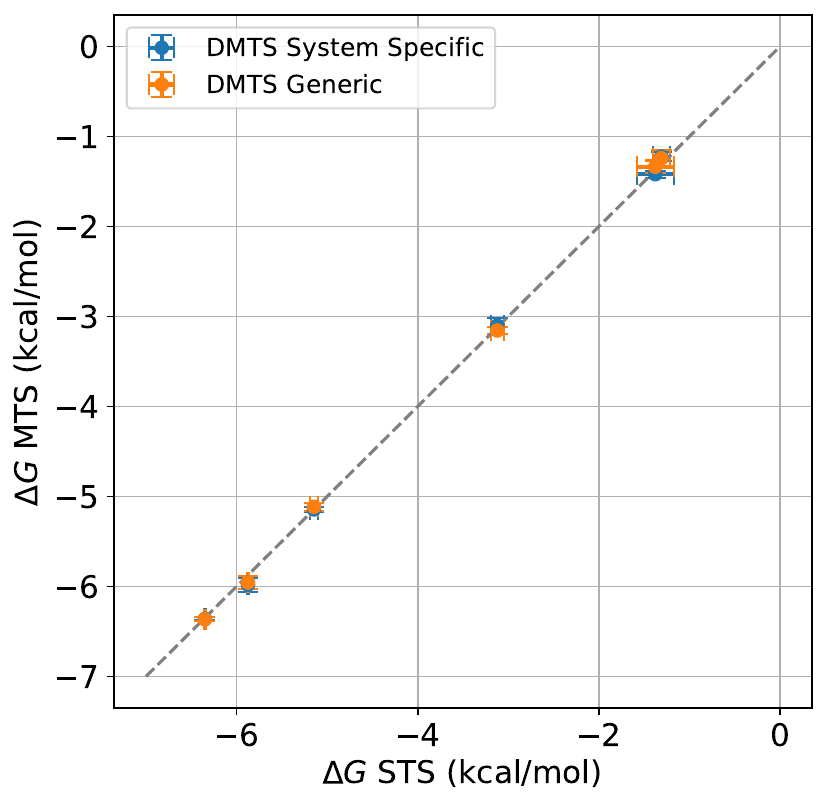}
    \caption{
        Hydration free energy of small molecules (water, ethanol, benzene, trimethylamine, diethylsulfide and acetic acid) using DMTS with system-specific model (blue points) and with small generic model (orange points) compared to the STS result.
        MAE is \SI{0.091}{kcal/mol} and \SI{0.103}{kcal/mol}, RMSE is \SI{0.124}{kcal/mol} and \SI{0.138}{kcal/mol}, R$^2$ is \SI{0.996}{} and \SI{0.995}{} for respectively the system-specific model and the generic model.
    }
    \label{fig:deltaG}
\end{figure}

Figure~\ref{fig:deltaG} shows the predicted HFEs compared to the STS reference values. The system-specific model achieved a mean absolute error (MAE) of \SI{0.091}{kcal/mol}, a root-mean-square error (RMSE) of \SI{0.124}{kcal/mol}, and a coefficient of determination ($R^2$) of \SI{0.996}{}. The generic foundation model also showed good performance, with an MAE of \SI{0.103}{kcal/mol}, an RMSE of \SI{0.138}{kcal/mol}, and $R^2$ of \SI{0.995}{}.

These results demonstrate that the DMTS scheme preserves high accuracy in free energy calculations while benefiting from a significant reduction in computational cost. They also show that both system-specific and generic models are capable of closely reproducing STS-level results.

\subsection{Protein-ligand complexes}

To validate our scheme on a biologically relevant system, we performed a \SI{20}{ns} NVT molecular dynamics simulation of the lysozyme–phenol complex (PDB ID: 4I7L) in explicit water. Because of the chemical diversity of this system, we do not train a system-specific model from scratch. We thus first assess the ability of the generic model to handle the system and then show that its performance can be improved via active-learning-driven fine tuning.

\begin{figure}
    \centering

    \begin{subfigure}{0.7\textwidth}
        \centering
        \includegraphics[width=\linewidth]{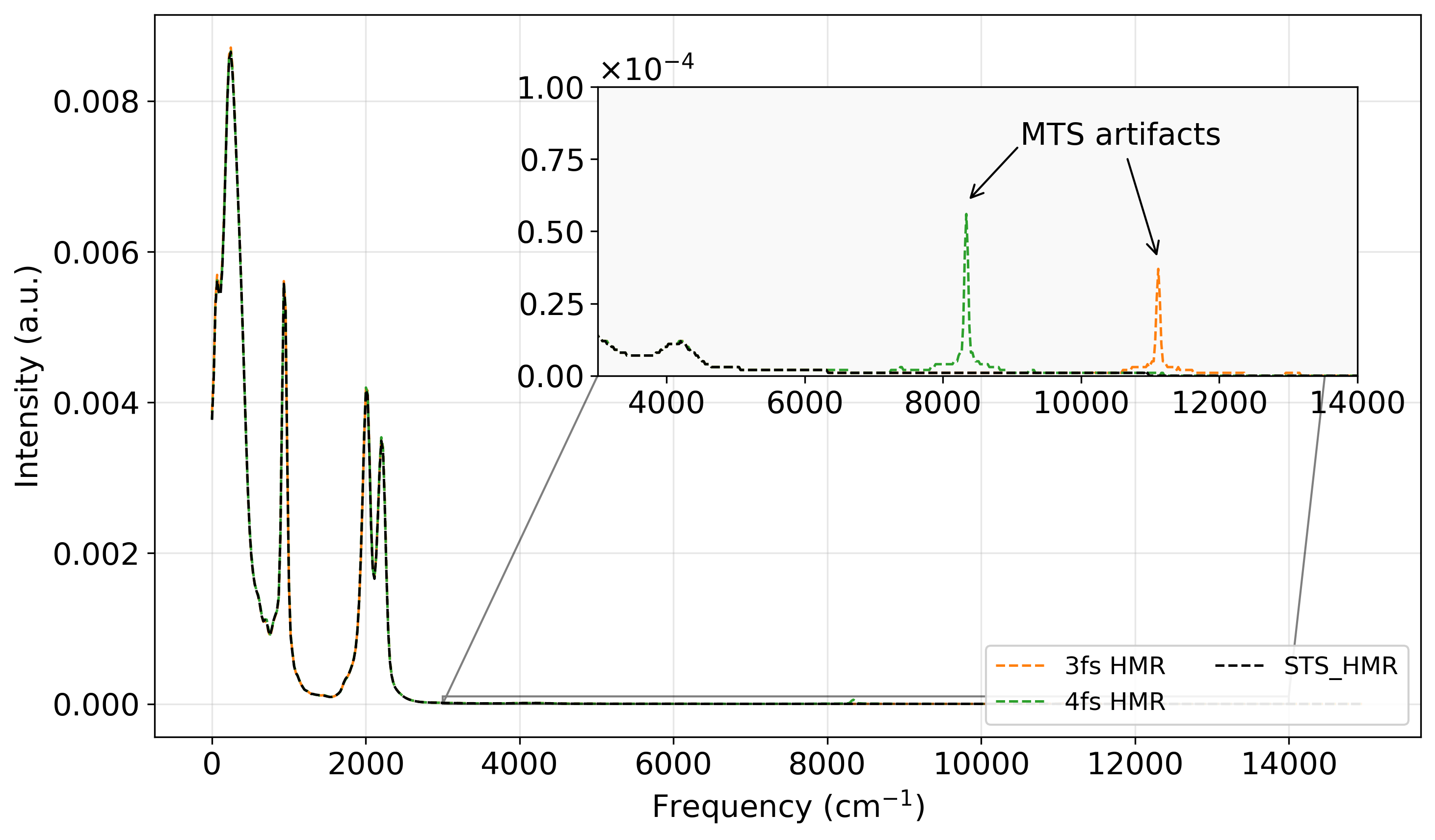}
        \caption{velocity autocorrelation spectrum}
        \label{fig:spectra_protein}
    \end{subfigure}
    
     \begin{subfigure}{0.7\textwidth}
        \centering
        \includegraphics[width=0.80\linewidth]{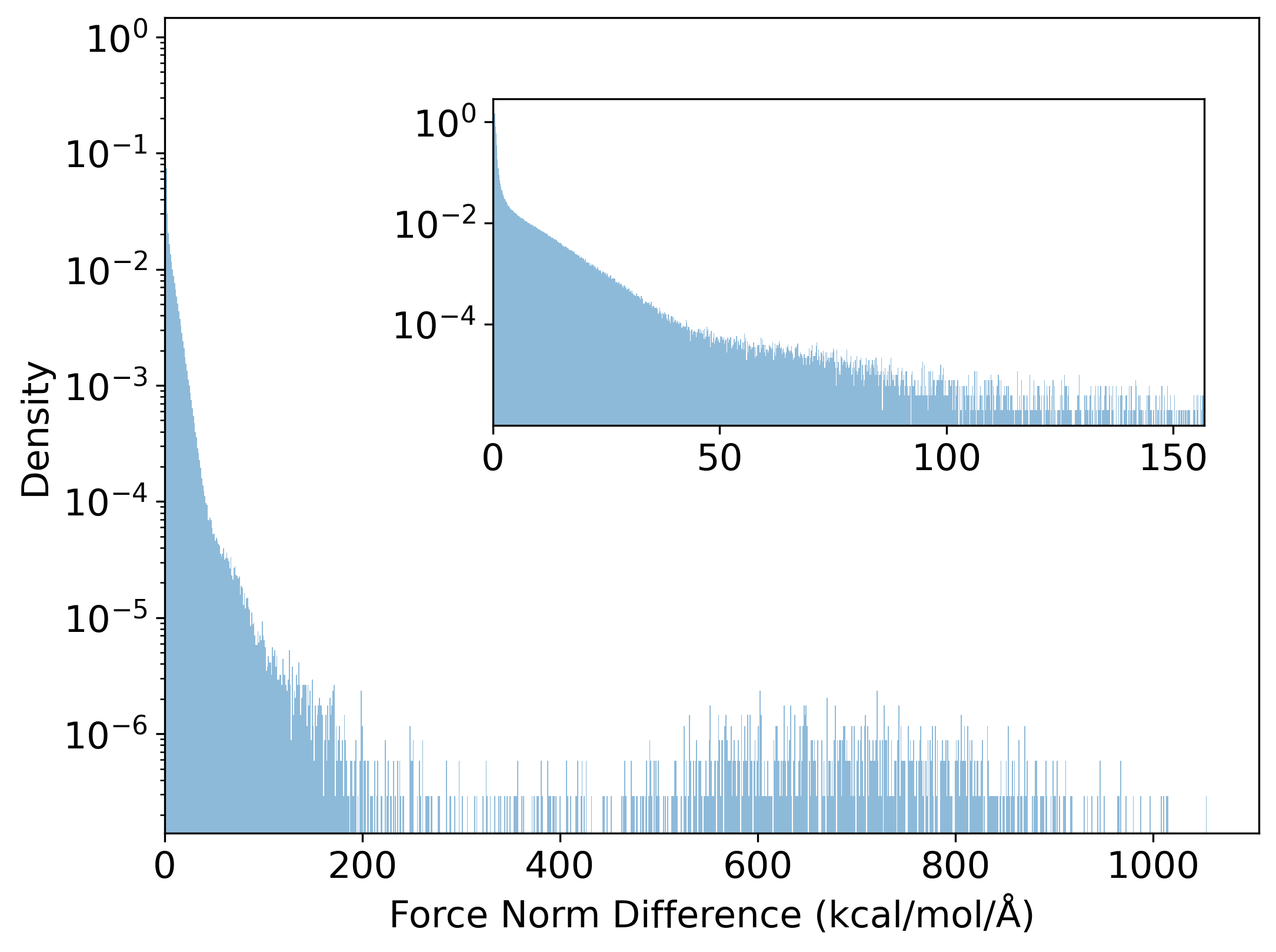}
        \caption{Distribution of the force difference}
        \label{fig:force_distribution}
    \end{subfigure}

    \caption{
        a) Average velocity autocorrelation spectra of hydrogen atoms for the phenol-lysozyme complex with HMR, comparing STS with MTS \SI{3}{fs} and \SI{4}{fs}.
        Insets zoom in high frequency regions containing MTS integration artifacts.\\
        b) Distribution (in log-scale) of the norm of the force differences for the generic model with respect to the FENNIX-Bio1(M) reference model. 
        The distributions are estimated over the first 500 frames of a phenol–lysozyme in water simulation.
    }
    \label{fig:spectra_comparison}
\end{figure}

\begin{figure}
    \centering
    \includegraphics[width=0.7\linewidth]{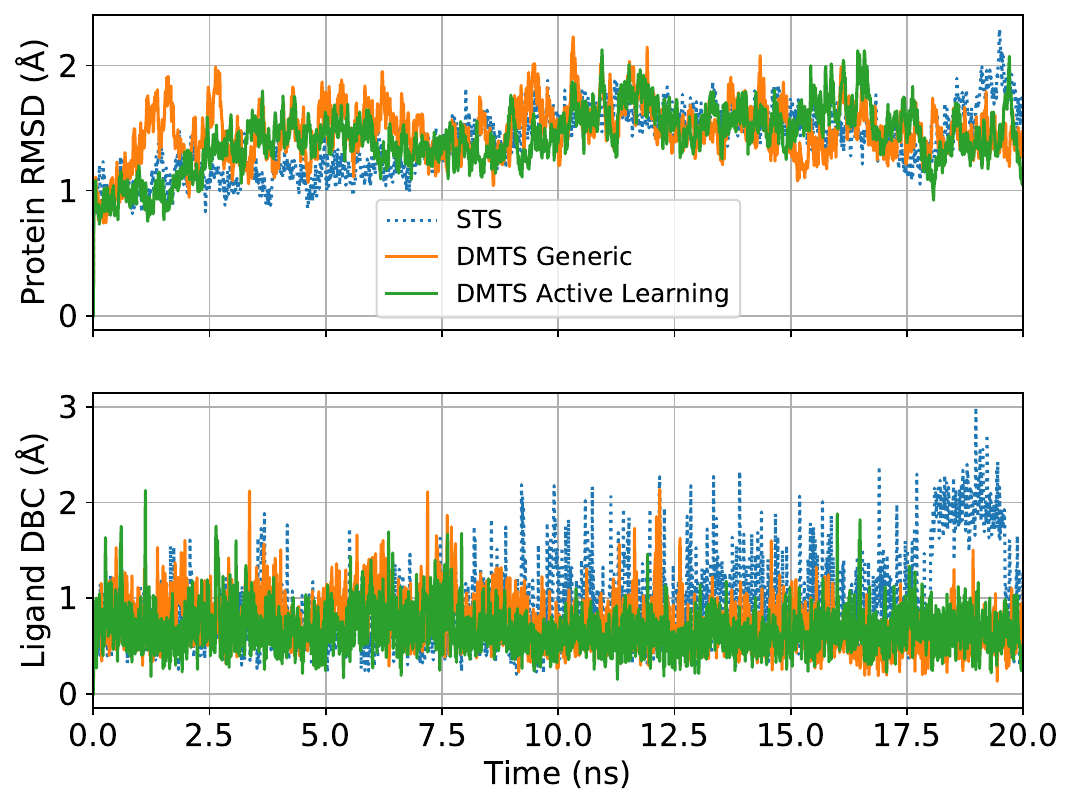}
    \caption{
        Time evolution of the protein backbone RMSD and the ligand’s Distance to Bound Configuration (DBC)\cite{salari2018streamlined} during a \SI{20}{ns} simulation of the lysozyme–phenol complex in water. Results obtained with the implemented MTS integrator (solid lines) are compared to those from a reference STS simulation (dotted lines). DMTS simulations use either the generic or active learned models combined with the FeNNix-Bio1(M) potential with an internal time step of \SI{1.75}{fs} and an external time step of \SI{3.5}{fs} with HMR for generic model and an internal-external time step of \SI{2}{fs}-\SI{4}{fs} with HMR for the active learned model.
    }
    \label{fig:protein_rmsd}
\end{figure}

Using an inner time step of \SI{1.75}{fs} and an outer time step of \SI{3.5}{fs} with HMR, we obtain stable simulations over \SI{20}{ns} with the generic model while preserving the protein's structure and the ligand's binding mode as shown in Figure~\ref{fig:protein_rmsd}. Note that simulations with a \SI{4}{fs} outer time step and HMR (which were stable for homogeneous systems) displayed instabilities after \SI{2}{ns}. The close similarity of the velocity autocorrelation spectra (Figure~\ref{fig:spectra_protein}) between STS and DMTS indicates that these instabilities are not due to the resonance phenomena described above.
They instead originate from the presence of ``holes'' in the potential energy surface of the small model, yielding large differences between its forces and the target ones. This is illustrated in Figure~\ref{fig:force_distribution} showing the distribution of the norm of the difference of forces between FeNNix-Bio1(M) and the distilled model along an STS trajectory of the protein-ligand system. We observe a dense distribution of values close to \SI{0}{kcal/mol/\angstrom} and a long-tailed distribution starting from \SI{150}{kcal/mol/\angstrom}. The latter is associated with large nonphysical force discrepancies yielding the infrequent instabilities that we observe in the DMTS simulation.

To address this limitation, we introduce two complementary strategies:

\textbf{Small message-passing model.} First, we explored enriching the small model architecture with explicit message passing. The intuition is that a more expressive model, with more parameters and deeper local information flow, should exhibit fewer ``holes'' in the potential energy surface, thereby reducing the occurrence of pathological configurations.
This strategy proved effective: with a single message passing layer, we obtained stable \SI{20}{ns} simulations using \SI{1}{fs}-\SI{4}{fs} inner-outer time steps, confirming the improved robustness of the small model (see Figure~S2 in Supporting information). However, the added architectural complexity significantly increases the inference cost, making the approach less attractive for large-scale production simulations (going from \SI{6.54}{ns/day} for the generic model to \SI{5.95}{ns/day} for the message passing model).

\textbf{Improving the small model via active learning.} We therefore investigated an alternative strategy that preserves the simplicity of the original model: active-learning-driven fine tuning. In this approach, we design an active-learning procedure that automatically detects frames and atoms where the force difference exceeds an unrealistic threshold of \SI{150}{kcal/mol/\angstrom}. For such frames, the integrator temporary reverts to the STS integrator and local clusters centered on the problematic atoms are added to an adaptive fine-tuning dataset. In total, curating the final dataset, required approximately \SI{400}{ps} of simulation (see details in Supporting Information). The generic small model is then refined on this curated dataset, systematically improving its robustness in regions of configuration space previously associated with instabilities.

Remarkably, this refinement procedure achieved, without increasing the model's complexity, stable \SI{20}{ns} simulations using up to \SI{2}{fs}-\SI{4}{fs} inner-outer time step.
The protein's structure and the ligand's binding mode were preserved as shown in Figure~\ref{fig:protein_rmsd}. Furthermore, we show in the Supporting Information (Figure~S1) that the potential energy is stable along the whole \SI{20}{ns} simulations in all cases.
With this configuration, we reach a production speed of \SI{7.45}{ns/day}, corresponding to a \SI{2.92}{} speed up compared to the STS simulations, strongly accelerating production simulations with biological systems. This demonstrates that targeted fine tuning can rival the stability gains of more expressive architectures while retaining a significantly lower inference cost. 
Looking ahead, this adaptive strategy could be further enhanced through more sophisticated active-learning schemes\cite{vanderoordHyperactiveLearningDatadriven2023, kulichenkoUncertaintydrivenDynamicsActive2023}.

%%%%%%%%%%%% Conclusion

We have introduced a practical and efficient Distilled Multiple Time Step (DMTS) scheme tailored to machine-learned force fields, enabling significant acceleration of molecular dynamics simulations without compromising accuracy for both static and dynamical properties. Our approach, based on a RESPA-like scheme, couples a fast, distilled neural network model with the accurate FeNNix-Bio1(M) reference potential. We demonstrated two complementary distillation strategies: (i) on-the-fly system-specific model distillation, and (ii) the use of a lightweight transferable model for broader applicability.

We achieve substantial speedups -- up to 4 in mostly homogeneous systems (bulk water and solvated small molecules) and 2.92 in large protein-ligand complexes -- while preserving key physical observables such as radial distribution functions, hydration free energies, diffusion coefficients, and protein-ligand structural properties. Note that this is a first estimation of the computational gains as the code has yet to be optimized further to handle the dual-level approach more efficiently.
All in all, our approach allows to reach above \SI{7}{ns} of simulation per day on a single A100 GPU for a realistic protein-ligand complex while preserving the \textit{ab initio}-like accuracy of the FeNNix-Bio1(M) model. In combination with well-established accelerated sampling schemes~\cite{Celerse2022,D1SC00145K,ansari2025lambda}, this enables truly large-scale simulations with foundation neural network potentials. 

Future work will focus on expanding the applicability of this approach through two main avenues. First, randomized time stepping could be explored to further increase the effective time step, while allowing for controlled velocity rescaling as proposed in stochastic RESPA variants, i.e. JUMP integrators~\cite{velocity_jumps}. 
Second, combining the fragment-based active-learning strategy with the generic message passing architecture offer a promising route forward. An active learning driven message passing model could selectively refine unstable regions on the fly, potentially enabling inner time step beyond \SI{4}{fs} while offsetting the higher inference cost of the message passing.

%%%%%%%%%%% Data availability
\section*{Data availability}
The generic pretrained model can be found on Github at \url{https://github.com/FeNNol-tools/FeNNol-PMC} 

\section*{Supporting Information}
Model's architecture and training procedure details.
Time evolution of the protein potential energy during \SI{20}{ns} simulations using STS and DMTS integration methods, generic and active learning models.  
Time evolution of the protein backbone RMSD and the ligand’s DBC during a \SI{20}{ns} simulation, using STS and DMTS integration methods, generic, active learning and message passing models. 

\section*{Acknowledgments}

This work has received funding from the European Research Council (ERC) under the European Union's Horizon 2020 research and innovation program (grant agreement No 810367), project EMC2 (JPP). Computations have been performed at IDRIS (Jean Zay) on GENCI Grants: no A0150712052 (J.-P. P.).

\bibliography{acs-achemso}

\newpage
\section*{TOC Graphic}
\begin{figure}
    \centering
    \includegraphics[width=0.9\linewidth]{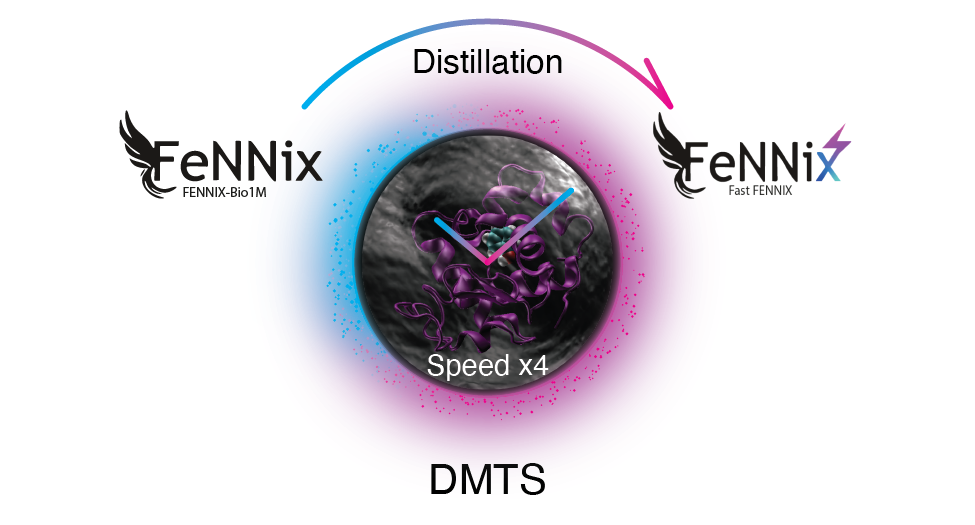}
    \caption*{For Table of Contents Only}
\end{figure}

\end{document}